\documentclass[sigconf]{acmart}
\renewcommand\footnotetextcopyrightpermission[1]{} % removes footnote with conference information in first column
\usepackage{latexsym}
\usepackage{hyperref}
\usepackage{relsize}
\usepackage{natbib}

\begin{document}
\title{Usability in the Larger Reality: A Contrarian Argument
for the Importance of Social and Political Considerations}
%\numberofauthors{1}
\author{
  Asad B. Sayeed
}
\affiliation{
	\institution{Department of Philosophy, Linguistics, and Theory of Science\\University of Gothenburg}
	\streetaddress{Box 200}
        \city{405 30 Gothenburg}
        \country{Sweden}
}

\email{asad.sayeed@gu.se}

%\raggedright
\begin{abstract}
Usability engineering is situated in a much larger social and institutional
context than is usually acknowledged by usability professionals in the
way that they define their field.  The definitions and processes used in
the improvement of user interfaces are subordinate to interests that
often have narrow goals, having adverse effects on user awareness and autonomy
that have further adverse effects on society as a whole.  These effects are
brought about by the way that knowledge about systems is limited by the
design of the interface that limits the user to tasks and goals defined
by organizations, often commercial ones.  

It is the point at which the structures of the user interface are defined
by usability professionals that sources of the limitation of knowledge 
can be identified.   These sources are defined by their reliance on a 
construction of the user's wants and needs that cyclically reinforce, through
the actual use of the interface, the user's own construction of her wants
and needs.  To alleviate this, it is necessary to come up with new processes
of user interface design that do not make assumptions about the user that
tend to subordinate the user, and it is also necessary to reconstruct the 
user as a participant in the interface.

This paper is intended as a philosophical critique, as described above, 
of the concept of usability engineering.  
It is also occasionally illustrated with examples of relevant situations 
involving the social and political effects of usability engineering.

{\it NOTE: This article was written in 2003 as a final project for a Master's course in human-computer interaction at the University of Ottawa. The author attempted to publish it at the time, but did not really understand the process. The ideas stand up fairly well, however, and he would like to contribute it as a comment on the current state of affairs. It is only lightly edited for format.}
\end{abstract}

\maketitle

\section{Introduction}
Over the years, work in various areas of science and engineering has become 
increasingly subject to public scrutiny over the effects it may have on
the environment, on individuals, and on society.  The public as a whole has 
begun to recognize that while the act of scientific observation and analysis 
itself may be value-neutral (subject to caveats of scientific ethics), the 
context in which science and engineering are conducted is often not.  Questions
arise about the benefits such research may bring and what harm.  These
questions are political ones and arise in debates about values.  It
thus behooves those who work in science and engineering to examine carefully
the ideological assumptions under which they choose to conduct research and 
how they intend for their work to be applied.

Genetic engineering is an obvious and recent case in point.  Questions about
biodiversity, effects on human health, economic effects, and so on have 
reached a point where researchers must pay attention and recognize the 
effect that their work has on society and the role of politics in their
research agenda---and that is just for agricultural genetic engineering.  
We have barely begun to scratch the surface of the moral, social, and
political implications of human germline engineering; it has everything
from medical implications to implications for class and social structure,
and it is thus a cynosure of public debate on science as well as the fuel
of any number of science-fiction novels.

What makes this science so politically significant?  Part of the answer is that
it either directly touches the lives of human beings or has future
implications that will.  This is the case, to a greater or lesser degree,
with other scientific and engineering research endeavours.  Usability 
engineering and the study thereof is a component of software engineering
that is explicitly devoted to studying the relationship between computer 
systems and users.  As such, it is one of the points at which the several
fields often known as information technology have
a more direct impact on people and society.  

Unfortunately, usability professionals and researchers appear often to discuss
deeply political and philosophical questions with only an eye toward 
narrow usability goals, seemingly not recognizing the relationship between 
their work and writing and larger questions of power and society.  For 
example, in a
recent panel discussion at CHI \citep{anderson}, Janice Rohn 
and Don Norman argued 
for a considerable period of time about how to gain influence within 
organizations by aligning themselves with powerful forces in the organization:
who are these powerful forces, what are their values, and so on.  
But the idea that this context of values may influence the direction
and goals of usability work does not appear to be a major factor in their
discussion; it is taken as a given that usability should primarily benefit 
the organization and not the individual user or the society in which the 
organization exists.  Indeed, Norman proposes that usability engineers
acquire business administration education, an education that is laden with
the values of management.  However, all is not lost: Rohn attempts to
attribute ``feature bloat'' in Microsoft products to Microsoft's monopolistic
control of the market; Norman, however, attempts to redirect blame to 
non-political factors.

Unstated or unexamined political ideas appear to surround or bracket 
discussions of usability in often quite blatant ways.  For instance,
the motto of the 2001 CHI conference was ``anyone. anywhere.''  
This statement appeared in bold text in the footer of papers published in the 
conference proceedings.  In this simple statement is packed a variety of 
potential meanings
\begin{itemize}
\item that anyone can apply usability techniques to any context.
\item that everyone should apply usability techniques to every relevant
context. 
\item that such things can benefit anyone, anywhere.
\item that usability research is already directed toward making technology
accessible to everyone.
\item that this is clearly a major goal.
\end{itemize}
Whether or not these implied claims are true is the matter at hand, though 
it will take far more than this report to support or deny all of them.
But it serves as further illustration the importance of context and
contextual assumptions in discussions of usability.

This is one example of well-known usability professionals either 
ignoring or merely scratching the surface of the political questions of
their claims.  However, there are writers in the field of usability who have
tried to direct attention to usability in the context of wider social goals.  
\citet{carroll}, for instance,
discuss the extension of usability evaluation techniques from 
the study of interactions between individual users of the systems in 
question to the integrated study of wide-ranging groups of people in 
their larger scale indirect interactions through the system.  They seek
to extend usability evaluation to handle the heterogeneous environments
that do not always reflect the relative uniformity of the traditional
business organizations.  They discuss the application of these techniques
to the evaluation of community networks, and they connect such evaluation
to the improvement of the social utility of computer systems, given 
an understanding that in some modern societies, there is a ``crisis of
community,'' wherein people's integration with local community life
is being stifled by the exigencies of the modern daily existence.  In
this case, it becomes important to see, among
other things, whether or not community networks
are being used in such a way that they serve community goals or whether
or not they are being used as on-ramps to other unrelated services. 

While it is clear that there is a growing trend in seeing usability
issues in social terms, however, there are still areas of 
socially significant usability assumptions that have not yet been 
questioned.  This report will deal with a small number of them.

\subsection{Objectives} \label{sec:obj}
The component of the debate about usability that I intend to focus on most
closely in this report is the politics of users' knowledge of the system 
and how it influences user interface design goals.  Specifically, I intend
to give a brief introduction to arguments
\begin{itemize} 
\item that definitions of usability are relative to the world-view of the 
usability professional, as partially explained above.
\item that this world-view assumes that users are static entities defined
by goals of organizations.
\item that this is connected to power relationships in human society.
\item that these relationships depend on limiting knowledge.
\item that expanding the the flow of knowledge to the user (indeed,
making more demands on the user) can have a subversive effect on power
relationships in society.
\end{itemize}
It is important to note that I am specifically acknowledging a goal here: the
empowerment of users to the detriment of interests that already have 
power over them.  It is not necessary for the reader to agree with the
way that this goal has been defined, but it should be noted that often
other writers' goals, as seen above, are not defined even if they exist.

\section{Usability and Goals}
I have spent the previous section introducing the relevance of social
and political goals to usability engineering.  However, in light of the
objectives above, it behooves me to narrow down the discussion to the 
goals against which I intend to evaluate aspects of contemporary usability
engineering.  After this, I will discuss how narrow definitions of usability
depend on the particularization of users and users' knowledge and 
its subjugation to organizational goals. Thereafter, I will discuss the
social impacts of user interfaces that narrow the information flow.

\subsection{Politically Motivated Goals} \label{sec:polgoals}
There are different kinds of goals, but in particular they can be classified
by their nature as general or situation-specific.  Carroll and Rosson,
for instance, define a set of social goals to inform the evaluation of 
a specific type of system created to alleviate a particular social problem
that they felt the need to address---community networks for solving the
crisis of community.  But \citet{muller}
do not have a specific type of system to evaluate.  They instead discuss
systems in a business context with the view that the values of business
cause usability analysis to see the relationship between humans and computers
primarily in terms of ``productivity and efficiency,'' a ``Tayloristic'' 
approach.  They would prefer to see an alternative characterization of
work expressed as the quality of three things: work product, work life, 
and communication.  They set quality of the work product against productivity,
since increasing productivity may lead to a decline in quality; they define 
quality of work life in terms of the intangible psychological
effects felt by the user; and they express the quality of communication in 
how computer systems affect the user's communications with other entities 
involved in the organization, such as customers or executives.

Only two of these three values are justified in terms of the user
or any human environment surrounding the user.  And unfortunately,
they still justify these criteria in terms of economic goals; for instance,
they justify the idea of designing user interfaces that improve the quality 
of work life in terms of the enthusiasm brought by the user to his or her 
work in the organization.  So this general view of making
use of ``non-Tayloristic'' values in user interface evaluation still has
the side effect of reinforcing a business goals, not social goals.

So 
\citet{carroll}
have a broad social view but a specific goal, while
\citet{muller},
starting with a radical hypothesis, ultimately lead to a business-oriented
view but one that applies to more general usability evaluation issues.  
But it would be better if there was a way to provide a broad social view
that informs usability analysis in general and not just for specific 
circumstances, finding the equator between these two poles. What sort of goals
are these?  As suggested in section \ref{sec:obj}, they are, among others,
goals relating to the distribution of power between individuals, a
distribution that can be affected and effected by the information that is 
provided by the user interface.  Then the goal becomes to design in
such a way
\begin{itemize}
\item that recognizes that power is enforced in organizations by
limiting certain kinds of knowledge.
\item that is based on recognizing the places in the design that limit such
knowledge.
\item that is willing to limit reliance on certain traditional
goals of user interface design, such as ``ease of use,'' that may often
depend on limiting knowledge to the end user.
\end{itemize}
On what sort of knowledge, then, is this goal predicated?  If it is simply 
the knowledge that must be imparted to the user in order to perform his
or her specific tasks---that is, domain knowledge---then there is no 
particular significance to these goals.  Under any reasonable set of
design goals, if domain knowledge is not being correctly imparted by
the user interface, then there is a problem.  No, in this case, the knowledge
in question is knowledge of how the system works, knowledge that would 
normally be confined to system designers, implementers, and expert users.

Why is it important for this knowledge to be exposed to end users?  Some of
this question will be answered in a later section (\ref{sec:limit}).  However,
to give a concrete example from consumer software, consider the growing
use of popular software on the Internet that sends data back to its 
manufacturer (KaZaa figures among these).  In this case, many users are
not aware what kind of information is being sent for data mining purposes.
It is likely that they are not aware that a program can monitor their
activities without their knowledge and transmit it without their explicit
actions.  As we move into a more paranoid security situation
with more government surveillance and scrutiny of everyday life, it is 
necessary that consumers become more aware of these things; designing 
other, more ethical Internet applications that do not hide so much of the
underlying Internet reality from end users may be a step in the direction
of user interfaces that promote social goals and social change.

Now, one may claim that this way of characterizing and justifying design
goals are simply relative to the person holding them and an entirely 
subjective manner.  Why should usability professionals be concerned about 
these goals?  The answer is to say, yes, these are relative to the views
of individuals, and there is indeed a subtext of social radicalism in
their construction.  However, this is a matter of degree; most people
would likely view at least the encouragement of user understanding of
security-related matters to be a positive effect.  But ultimately, the
actual moral value of such goals goes into philosophy and beyond the scope
of this report.

\subsection{Usability and Roles} \label{sec:roles}
How is system knowledge apportioned and limited?  How do current 
concepts of usability engineering help to encourage this
state of affairs?  This is inextricably tied up in how usability 
professionals see themselves and their field, as with Rohn and Norman
are described to do in the interview mentioned in introduction to this 
document \citep{anderson}.
I will therefore discuss a definition of usability and usability engineering
as provided by Butler, who seeks to describe usability engineering's past
accomplishments and rise to importance
\citep{butler}.

To begin with, \citet{butler} defines usability
to be the ``effectiveness of interaction between human operators and their 
machines.''  Butler attempts to define this in objective terms by claiming
that this effectiveness reflects ``how well intended users can master and 
perform tasks on the system,'' given empirical measurement brought to bear 
by the observation of ``actual user performance of frequent and critical 
tasks on the system.''  Usability engineering, therefore, is a ``new 
discipline'' that serves ``to address system usability in a reliable
and replicable manner.'' 

It is important to note that Butler provides this definition only 
after he provides background information about the larger context of
usability in much the same way that I have arrived at this point of definition
after describing the larger context in which I intend this discussion.
However, he is not at all self-conscious about the nature of his background.
He specifies his primary motive in the idea of the success of a 
system, clearly defined in terms of its commercial value on the marketplace
or its cost to a buyer.  

How does poor usability cost the buyer?  According to
\citet{butler}, it can cause businesses to lose opportunities 
to ``re-engineer their processes.''  The work that a user does is defined
in terms of the institution; the value of the user interface is entirely
contingent on how it allows the user to serve the institution, and, more
abstractly, how it allows the institution to manipulate the environment
of the user, as represented by the processes of the business organization.
Indeed, ``work pressures'' give the user less time to learn the interface,
making usability crucial to efficiency and reducing overhead.  

Task analysis is one of the main tools which Butler expects usability
professionals to use in order to provide information as to how the 
interface should be designed.  But these tasks are apportioned by the
organization.  In fact, it is the tasks that apportion users into different
roles in the organization---indeed, that is one way that organizational
roles can be defined: by the tasks they imply.
In other words, usability as seen by Butler, whose article
\citep{butler} is devoted to discussing recent trends, is not only defined 
by the tasks that the user must perform, but it also partially circumscribes
the user's role.  It is designed, by its reliance on task analysis to 
improve business efficiency, to reinforce aspects of society that
are by their very nature political and philosophical.

By what mechanism does usability come to define the role of the user?  Butler
notes that an important part of usability work is determining what 
should be automated by the machine and what should be performed directly
by the user.  This is one direct way of limiting the user's role.  But
what about the interface itself?  Does it contain some characteristic that
constrains the user and helps to subordinate the user to the goals 
of the institution?  Part of the answer to this is the way in which
user interfaces constrain the access of the user to knowledge 
about the system, an issue to which I have alluded
in the previous section and to which I will return in section \ref{sec:limit}. 
Furthermore, it is the presupposition of the organizational roots of tasks
that cause usability analyses to prescribe user interfaces that 
promote these constraints.

\subsection{Effects of Limiting Knowledge} \label{sec:limit}
By now I have reached a point where I have described usability as a force 
that affects the user's power the define his or her own role in the 
organization and in society---in other words, notions of usability can
constrain the user's knowledge and therefore the user's power.  However,
I have not as yet described some of the means by which this is accomplished.

Much of the answer is discussed by \citet{johnson-eilola}.  
He is concerned most directly with training and documentation aspects
of computer systems, but he also applies his ideas to the philosophical
underpinnings of user interface design as well.  

Johnson-Eilola notes that as users have come to expect that computer 
technology becomes increasingly automated and ``anticipates our every desire,''
it becomes more difficult---``pointless''---for users to ``think
critically about the operation of the machine and our position within it.'' 
Indeed, users become less aware of how technologies ``construct positions
that users assume'' the more invisible the workings of technology seem to be.
In other words, users come to expect that technology will automatically satisfy
their needs, requiring less and less participation on their part.

Designers of software respond to these expectations by designing is
such a way that ``users shouldn't have to think.''  But this
creates a vicious cycle, as it reinforces that very expectation.  As users
come to expect that their contribution in terms of effort and thought 
will be reduced, users become less and less capable of defining the
reasons for what they expect.  John\-son-Eilola refers to this situation
as the ``politics of amnesia.'' 

Is amnesia always bad?  Should the user be involved in every detail of
the operation of the system?  Where is automation-induced amnesia a
good thing and where is it bad?  John\-son-Eilola notes that modern society
requires such a complex flow of information that it is impossible that 
human beings be required to process all of it.  But the cycle that 
causes designers of user interfaces and writers of documentation to reduce
the flow of information to the user tends to a limit where the user is
deprived of any information that would allow the construction of a
mental model of the system.  ``Amnesia'' is not always a bad thing, 
but Johnson-Eilola wants to identify the ways in which user interfaces
go too far rather than help users to ``actively build memories and
experiences.''

In section \ref{sec:polgoals}, I briefly gave an example of situations
in which the user's lack of a mental model of Internet communication
leads to situations in which the user's data can be exploited by 
common software that communicates without the user's consent.  I gave
the example of the music-finding software, KaZaa, and this piece of 
software illustrates this point quite well.  KaZaa is a program whose 
purpose it is to automate as much as possible the user's search for music
and other files; not only does its automation exploit the user's data, it 
itself reinforces the patterns that give rise to the user's lack of knowledge. 

But KaZaa is a minor example of the unknowing loss of autonomy experienced
by the user.  A much larger example is that of email standards and the
marketplace.  This is discussed in detail by \citet{jakobs} in 
their paper on the relationship between users and the standardization process
that focuses on the reluctance of organizations (here apparently defined 
as a sufficient representative of the user) to become involved in standards
bodies.  While Jakobs et al. describe a number of reasons for this phenomenon
with respect to email standards, one of them is that the organizations and 
the people responsible for setting up email therein are used to viewing 
email through the products and services being used to provide it.  

This necessarily makes the user interface of these products the main arbiter of
how email and email standards are viewed by the end user.  Since leading 
commercial email packages such as Microsoft Exchange
tend to obscure the challenges that are faced in other, 
heterogeneous environments where multiple email packages are used, the user 
organization becomes increasingly less aware of the issues of 
interoperability that standards are intended to solve.  This reinforces
reliance on proprietary products, which in turn reinforces the lack of 
awareness, in a similar cycle as the politics of amnesia.  As such products
as Microsoft Exchange are generally integrated with other Microsoft products,
it further increases the prevalence of Microsoft in the organization, leading
to the near-monopoly that Microsoft presently has in the 
marketplace\footnote{I take it here for granted that monopolies are bad.  
The negative economic, social, and technological impact of monopolies 
are beyond the scope of this report.}.  This demonstrates how user
knowledge and the politics of amnesia has an effect on user choices and
the technological situation. 

In a sense, the subordination of the concept of usability to institutional 
goals creates a series of interlocking amnesiac cycles of various scales
(from KaZaa to monopolies)
that inhibit a practice of usability that realizes the goal of returning
autonomy to the user.  The remaining task is to locate the conjunction
of these cycles, the point which they all have in
common, in order to diminish their effects.

\section{Users, Design, and Knowledge}
I have accomplished the main task of discussing some ways in which 
usability engineering and the goals and assumptions 
behind it control knowledge in such a way that the user is inadvertently 
subordinated to institutional goals.  But there is the remaining
task of explaining the aspects of user interface design that wherein
the control of knowledge is actually implemented.  

The fundamental link in the chain of 
stifling invisibility are the design practices followed by usability
professionals in the way that they choose to arrange and structure
user interfaces;  while these practices are informed by their unacknowledged
political understandings, they could have no understanding or reinforcement
without a point at which they become real.  Task analysis and other
organizationally embedded processes do not have any relevant 
effect on reality until the work of building the user interface is 
accomplished.

Consequently, in the remainder of this section, I will discuss the flaws 
with the philosophy of user interface design that appear to be prevalent 
in usability engineering; I will discuss a potential way of alleviating
these flaws through by re-orienting established design heuristics; and 
I will expand the discussion not only to the interface itself, but to
reconstructing the user as a participant in the interface.

\subsection{Invisibility and Design} \label{sec:invis}
\citet{johnson-eilola} goes on to discuss the aspects of
the politics of amnesia that cause these effects as they are expressed
in documentation and user interfaces.  He notes that the 
most important technologies that human beings use are 
the most invisible and used without thought: ``It's hard to argue with
something that's not there.''  But when it comes to user interfaces, the
invisibility that defeats user awareness is often a design choice.  
According to Johnson-Eilola, designers and documenters operate on
``Shannon and Weaver's'' communication model from the 1940s.  In this model, 
the sender imparts information to a receiver through a limited channel;
the task of the receiver is simply to present itself.  Very few 
models of communication involving human beings rely on this view of the
human receiver, having instead complex models of the receiver's response, 
interpretation, social situation, and so on.  But the fact that this model
is simple for designers to conceive and happens generally to ``work well 
enough'' to help most users to accomplish their desired tasks allows this
model to persist. 

The first component of the expression of the Shannon-Weaver model in 
user interface design is summed up by Johnson-Eilola as ``Do as I say, not
as I do.''  In Johnson-Eilola's specific area of computer documentation, this
aspect of invisibility is expressed in the way that documentation writers 
measure the success of their texts: the ease at which the users are able to
follow the instructions.  But most documentation writers are also aware that
users interact with texts in an often highly creative process, by doing
such things as linking the information in the texts to prior experiences.  
So users are paradoxically encouraged to view the text as trivial while
the information in the texts become more complex with the technology.  

I can easily extrapolate this to user interface design beyond 
documentation.  The invisibility of technology as produced by automation is a 
result of designing the interface to satisfy the user's desire 
for simplicity while enabling the user to perform complex actions in 
such a way that telescopes complexity into user actions that minimize 
the user involvement.  Though invisibility is not always bad, it is the 
assumption that this paradox must always be satisfied in favour of simplicity
that leads to design decisions that produce invisibility that reduces
the ability of the user to imagine a social criticism.

This telescoping of complexity leads into the second component of the 
denial of the receiver/user's agency produced by a Shannonistic model.  
Once again, to put it in Johnson-Eilola's pithy wording \citep{johnson-eilola},
``real learning disappears in the collapse of time.''  Johnson-Eilola 
emphasizes the difference between learning and being trained.  Learning
implies an attempt at increasing the possibility of future creative use
of the acquired knowledge; but merely to be trained is to acquire knowledge
in such a way that it is only useful for a particular task.  Johnson-Eilola
uses the example of a hammer: should the use of the hammer be taught only
sufficiently to succeed in the task of banging nails, or should it be taught
in such a way that allows the hammer's user to acquire carpentry skills
in the future?  To Johnson-Eilola, the latter is 
clearly the more desirable option---it is the true learning.

In the desire to improve efficiency, whose nature with respect to user 
interfaces has been discussed in previous sections, the time required
for documentation to impart knowledge is often reduced to the time required
for training, not learning \citep{johnson-eilola}.  Once again, this can
be extrapolated to user interface design.  Aspects of the 
user interface that can be designed to teach (that is, bring about
real learning) are instead co-opted into performing tasks in an invisible 
manner.  Johnson-Eilola brings up the example of ``wizards,'' tools that
can be used to allow the system to engage in complex interactions with the
user.  In their potential to expand the user's understanding of the system
by revealing through interactions what the system is accomplishing, instead
wizards are typically given the role of automating for the user complex tasks
that do not lend themselves well to this sort of automation.  Microsoft
Word, for instance, contains many wizards for creating different kinds of 
documents, but these wizards do not discuss well how and why these documents
are structured so that learning---and therefore user autonomy---can ensue.

A last issue brought up by Johnson-Eilola that can be translated into this
discussion is that of the ``collapse of critical distance'' brought about
by the expectations of speed: 
``At the speed of light, time ceases to be an issue.''  From the
point of view that I am presenting, it is impossible
for users to regain autonomy and power, to reduce the negative effects
of the politics of invisibility and amnesia, if their capacity for 
criticizing the system's wider effects is diminished.  The time expectations
of the user, that ``the distance between desire and result should be zero'' 
\citep{johnson-eilola}, are cyclically reinforced to produce a situation
wherein the user does not even have the opportunity to criticize.  
Johnson-Eilola discusses the tendency of word processing and other productivity
tools to move important interface elements from invisible places such as menus
to immediately visible places such as tool-bars.  It may seem paradoxical,
but this kind of change from apparent invisibility to visibility---reducing 
the need for the user to search through menus---actually serves to increase
the overall invisibility of the context of actions and the whole menu 
arrangement of action, decreasing the critical distance of the user.  

I have thus used Johnson-Eilola's ideas to characterize the nature and
effects of three components of the Shannonistic approach to user interface 
design: 
\begin{itemize}
\item Designing so that complex tasks are performed in deceptively
simple ways destroys the possibility of creative and critical interaction
with the system.
\item Focusing design on specific tasks reduces the opportunity for real
learning.
\item In making actions more easily available to the user, there is a
price to be paid in terms of encouraging user assumptions and reducing
the opportunity for thought.
\end{itemize}
As these issues are in conflict with the goals laid out in section 
\ref{sec:polgoals} and serve, in the long term, to subordinate the 
user in ways discussed in sections \ref{sec:roles} and 
\ref{sec:limit}, we need to revise the approach to user interface design
in order to diminish the effects of a Shannonistic model.

\subsection{Beneficial Complexity}
The primary purpose of this report has been from the beginning to discuss
the concept of usability from a political and philosophical standpoint.  
However, I shall now briefly turn to what is an even more complicated 
discussion of what to do about the critique, relating it to ideas that already
exist in the usability literature but were not necessarily intended to be 
used to implement a philosophical programme.  

There are a number of techniques used to design user interfaces, but one 
particularly interesting one is the use of design heuristics.  Heuristics 
attempt to present a manageable set of guidelines that capture good practices
in user interface design; they can be used both in prototyping and 
in evaluation.  Heuristics present a boiled-down encoding of how user
interface professionals view the structure of the user interface, and 
thus they present a useful target for reform.  From such a change, there can
be a re-evaluation of the larger construction of the concept of usability.

An interesting proposal for altering the way we see user interfaces 
comes from \citet{gentner}.  They discuss
the history of user interface design guidelines as they have supported the
WIMP (windows, icons, menus, pointers) model with the view that 
the reliance on certain guidelines has stymied the production of new 
user interface paradigms beyond WIMP.  In effect, their concern is
not social like the motivation behind this article; nevertheless, there is
potential in their proposal for a politically-aware practice of usability
engineering.

For \citet{gentner}, an important exemplar of modern
user interface design guidelines are those for the Macintosh, published by
Apple in the 1992 (though they appear to be much older).  These design 
guidelines were set under certain assumptions, some of which involved the
need for the guidelines to help Apple to sell products to users.  Gentner
and Nielsen wish to perform a thought experiment: what if user interface
design was performed under heuristics that violated those guidelines?  They
call these the ``Anti-Mac'' heuristics and proceed to define and contrast 
them with the Mac interface guidelines.  Table \ref{tab:antimac} lists
these contrasted principles by name.
\begin{table}
\centering
\begin{tabular}{|c|c|}
\hline
\textbf{Mac} & \textbf{Anti-Mac} \\
\hline\hline
Metaphors & Reality \\
\hline
Direct Manipulation & Delegation \\
\hline
See and Point & Describe and Command \\
\hline
Consistency & Diversity \\
\hline
WYSIWYG & Represent Meaning \\
\hline
User Control & Shared Control \\
\hline
Feedback and Dialog & System Handles Details \\
\hline
Forgiveness & Model User Actions \\
\hline
Aesthetic Integrity & Graphic Variety \\
\hline 
Modelessness & Richer Cues \\
\hline
\end{tabular}
\caption{Mac vs. Anti-Mac design principles;  
after Gentner and Nielsen.} \label{tab:antimac}
\end{table}

Some of these principles bear directly on the matter at hand.  The
Mac principle of See and Point, wherein the user is limited to clicking
on only the presented objects, represents a form of limitation on the
user that coincides with invisibility; the user can do nothing but
what is presented there, which may be very limited in relation to what
the user may eventually wish to do.  The Anti-Mac principle of Describe 
and Command, 
however, represents the view that language is central, that subordinating
the user to a limited, ``easy,'' interface is a step back in human
communications; in the Anti-Mac interface, Gentner and Nielsen suggest,
the user can instead compose commands in a simple language (so as not
to set the infeasible requirement that computers understand 
the entirety of human language) that express a much broader range of
interactions than See and Point.  The use of language can help to return 
control over complex actions to the user, rather than requiring that
the complex actions be inscribed in the interface itself.  

Another contrast between the Mac design guidelines and Gentner and 
Nielsen's Anti-Mac guidelines is the role of WYSIWYG.  WYSIWYG stands
for the well-known dictum, ``What You See Is What You Get,'' so that
what is displayed represents as closely as possible what needs to be 
produced after the user performs her tasks (such as an image of
a printed page in a word processor).   But as \citet{gentner} point out, 
often this merely translates to 
``What You See Is All You Get.''  The Mac guidelines specify WYSIWYG; 
the Anti-Mac guidelines focus on representing the meaning of 
interface objects that appear to the user.  Indeed, Gentner and Nielsen 
call for Anti-Mac systems to have richer internal and external 
representations of 
objects with which the user interacts.  Richer internal representations
are needed because the freedom of interaction produced by a language-oriented
user interface requires more complex information about the objects.  A richer
external representation with more complex ways of interacting with objects
allows the user greater autonomy and freedom in using the objects.  
From the point of view of autonomy as a social good, richer internal 
and external representations of user objects allow users to expand their
use beyond the tasks intended by the designers of the interface, reducing
their subordination to the interests that establish the requirements
of the design.

Not all is well, however, with the Anti-Mac guidelines.  A Mac design
guideline is Feedback and Dialog.  Contrasted to this is the 
Anti-Mac guideline of System Handles Details.  Defined any reasonable
way, letting the system handle the details of an action promotes a
form of invisibility-inducing automation.  The Mac guidelines, however,
are more inclusive of the user as a participant in the process.
The situation can be rectified somewhat by letting the system handle the
details in such a way that the user remains informed about the actions
that the system is taking, as well as allowing the user control over
the nature of the automation itself---once again implying a need for
the expressive power of language in scripting or some other medium.  
Nevertheless, while containing some aspects that further an agenda of user
power and autonomy, the priorities of Gentner and Nielsen are considerably
more traditional.

\citet{gentner} mention that ultimately the Anti-Mac
interface is designed for users with a greater level of expertise than 
the Mac interface.  Considering the current levels of computer
use by young people, they assume that in the future, a greater and
greater proportion of the population will be classed as expert users.  Then
the trade-off between ease-of-use and power will be much easier to make.

But to assume that this will be the case also assumes that expertise will
be generated without social change.  In order for users to regain 
autonomy and critical distance, it is not enough that more complex
user interface paradigms be implemented passively over time.  This 
passive mode envisioned by Gentner and Nielsen does not optimally raise
the consciousness of the user to a state of autonomy from the parameters
of the institutionally-oriented task analysis performed in the user 
interface design.  Instead, an active mode of creating expertise is required,
a user interface design paradigm that does not merely allow the user
to take advantage of powerful features, but rather encourages and pushes the
user into a heightened awareness of the capabilities of the system. 
(The benefit of this is explored in section \ref{sec:expert}.)  

Already there are interfaces that display some of the Anti-Mac 
characteristics: the user interfaces for current Unix and Linux systems.  
The extensive reliance on command prompts even in graphical situations
takes advantage of the power of language, albeit perhaps in a less forgiving
way than that envisioned by Gentner and Nielsen.  And it may very well be
a less forgiving user interface that is necessary to create user growth
and empowerment, a user interface that reflects more closely the 
underlying state of the system and thus the possibilities 
for autonomy that emerge therefrom.  

\subsection{Expertise and Empowerment} \label{sec:expert}
Mention was made in section \ref{sec:invis} of the practice of indulging 
the user's own desire for simplicity and efficiency at the expense of 
critical distance and autonomy; it has also been noted that 
focusing on satisfying this desire {\it creates} this desire, as the ability
to learn and to criticise becomes increasingly less possible.  So
we must not only begin to reformulate the heuristics of user interface 
design, which is one half the equation, but also how we deal with users.

Furthermore, an agenda of beneficial complexification runs the risk of 
reversing some positive social gains made by simplified user interfaces, in 
particular the accessibility of the user interfaces to disadvantaged groups
such as women and minorities.  Rather than sacrifice the goal of increasing
user involvement in computer systems, we should rather seek to understand why 
it is that these groups are excluded from technology in the first place.  

It is well known that for most of recorded history, technology has been defined
ultimately as a masculine space.  Technological spaces have generally
excluded women; women are still vastly in the minority  in computer science 
and  electrical engineering, among other ``technology-heavy'' fields.
In his discussion of an attempt to increase women's involvement in technology
\citep{herman}, Herman notes that some radical feminists have taken the 
masculinity of technological spaces as fundamental to technology, seeing,
among other things, how it has generally been very closely connected to 
such historically masculine institutions as the military-industrial complex.
In a sense, if one takes this viewpoint to its limit, opening technological
spaces to women can often be constructed as a masculine intrusion on women's 
spaces, supplanting a transgressive act.  

This viewpoint also ignores the fact that there have been technologies that
also have liberated women---not perfectly, but they have allowed women
to resist patriarchal institutions.  There is a kind of technological dialectic
whereby, indeed, women's technological gains can be co-opted.  But in this
technological world, it is necessary that women be able to reclaim 
technological spaces, if only for the fact that both men's and women's lives
are deeply affected by it.  Indeed, such a transgression is an essential
part of the critique of the social effects of technology.

\citet{herman} describes an organization in Manchester, England,
called the Women's Electronic Virtual Hall (WEVH).  The WEVH was part of
a wider set of Electronic Virtual Halls (EVH) intended to make
technology more accessible to disadvantaged groups.  One of the successes
of the WEVH, as described by Herman, is the introduction of women 
participants to technology in a direct and hands-on way, not just as a tool
but as an item of interest in itself.  Courses were offered in which women
without prior technical experience were encouraged to touch technology, to be 
involved with technology not just as a transgressive act but as a normal
experience; these courses involved training women in such things as 
assembling the internal hardware of computers (motherboards and so on), 
plugging together PCs, setting up networks, trouble-shooting, and 
maintenance---not just using the computer, but being involved in it.  

It is the promotion of expertise that allows the completion of the other
side of the usability equation as discussed in this paper---not just expecting 
that users be passive consumers of computer technology, wholly determined and 
constructed by their experience of the user interface, but rather that, as 
active agents, they too are a partner in the construction of the usability.  
Expecting nothing from the users but the completion of their tasks is 
precisely the self-reinforcing Tayloristic goal that usability engineers 
seem not to question, as I have discussed near the beginning of this 
paper---it is associated with the ubiquitous obsession with productivity and 
efficiency, not autonomy and creativity.  Women who participated in 
the WEVH are now better empowered to participate in the technological world
not merely as users, but as critical thinkers; they are more likely to migrate
from being users who are constructed by usability engineers to being users
aware of their own construction and the construction of technology as an 
influence on society.

Ultimately, invisibility must be challenged both from the design
of the interface and from the perspective of the user.

\section{Conclusion}
In section \ref{sec:obj}, I discussed my main goal in this paper which was
to provide brief introductions to five political and philosophical
ideas regarding usability, knowledge and power.  In the following sections,
I proceeded to fulfill each of these objectives in some measure.  I discussed
a mainstream definition of usability in the context of current work.  I 
discussed the goals of usability engineering in order to lead into a
discussion expanding these goals to a practice that serves to challenge
the social order rather than serving the social order.  I gave examples of 
both subtle and blatant exertions of power between varying interests in 
society, and I connected these to the way in which knowledge flows between 
these interests.  And finally, I discussed the ways in which this flow of
knowledge is related to design practices and to the view of the user
held by usability professionals and potential ways of subverting 
these.

When we subordinate usability to the goals of institutions and the short term
goals of users, we actually persist in reinforcing a false consciousness 
of power that is actually invested in Tayloristic goals of productivity 
and efficiency, concepts that do not liberate the user but instead subordinate
the user to institutional goals as well; if usability is subordinated, then 
so the user will be.  

Often certain programmers and engineers refer, not quite seriously, to 
end-users as ``stupid.''  While this appears to be a rejected perspective
in usability engineering, it may be time to rethink that.  Perhaps users
really {\it are} stupid, stupid in that their level of understanding prevents
them from obtaining a level of autonomy from the systems they use, but
rather they are subordinated to an institutional idea of tasks and goals. 
And perhaps this is partly because users have been constructed that way by
those who design user interfaces.  Perhaps making a more difficult 
user interface is not so bad; perhaps ``ease-of-use'' is not an end in itself.
In any case, it is time to reevaluate how usability professionals affect the 
world through their paradigms and decisions.

\subsection{Practicality}
In several previous sections, I have mentioned examples of where
the existing concepts of usability engineering serve to subordinate 
the user and points at which this control can be broken.  But an 
important remaining question is how the usability practitioner can actually
take advantage of these points if he or she is so inclined.  A major 
topic I have avoided dealing with so far is the reward system by which 
usability specialists are formed into a discipline.  Usability professionals
are ultimately hired to perform tasks that their employers and clients
require them to do and expect them to perform within the material and
ideological bounds that benefit them.  A subversive practice of 
usability engineering in which efficiency and productivity are not given
primacy may ultimately be impossible for usability professionals 
to implement in real life in a large, obvious, and direct scale.

Nevertheless, the very act of discussing such a subversive practice in
an environment where the institutional goals of usability engineering 
are assumed is itself useful and important.  It transgresses against the
boundaries of usability discourse that are already set.  Eventually, it
may result in small acts of user empowerment that may grow over time.

Perhaps a subversive usability practice can exist.  Perhaps it exists already
in the freewheeling development of freeware tools.  We shall see.

\subsection{Subversion}
So all this leaves a final question: what research directions to take 
from this point?  At the moment, the way to a truly transgressive practice
of usability engineering is not clear at all.  While a new way of looking at
user interface design heuristics and a new way of looking at the
users themselves are very well and good, there is still yet another
component that needs to be inserted into the equation: the usability 
engineers.  We need a new way of looking at usability engineering as
a discipline.  For instance, when they are not merely copying the interfaces
of popular Microsoft Windows products, the freewheeling processes that underlie
the building of user interfaces for Unix/Linux-based open source freeware 
appear to produce somewhat ``messier'' user interfaces, but ones in which
the user often has far more control about the nature of her work and her
interactions with the system.  It is as though the undisciplined processes of 
open source software design by their very nature produce interfaces more
conducive to user learning and to user autonomy.  Further work is therefore
required to discuss whether or not open source interfaces really are 
more conducive to end user autonomy, in what ways this is so, and what specific
characteristics of open source design processes tend to produce these 
effects.

If usability engineering is ever to achieve a point where a practice that 
helps to subvert existing power relationships in society can emerge, then 
it is necessary to investigate power relationships involved in the current
practice of usability engineering.  Let the user no longer be used.

\section*{Acknowledgements}
Thanks to Dr. T. Lethbridge at the University 
of Ottawa for comments on the general direction of the paper.
Thanks also to Mr. C. McFarlane at Carleton University
for his comments on some of the political and philosophical aspects
of this paper.  This work was partially supported by the Natural Sciences
and Engineering Research Council of Canada.

\bibliographystyle{apalike}
\bibliography{indepth}

%\balancecolumns

\end{document}